\documentclass[a4paper, UKenglish, cleveref, thm-restate]{lipics-v2021}

\pdfoutput=1 
\hideLIPIcs  

\theoremstyle{definition}
\newtheorem{algorithm}{Algorithm}[section]

\graphicspath{{./graphics/}}

\title{Local Routing on Ordered $\Theta$-Graphs}

\author{André {van Renssen}}{The University of Sydney, Australia}{andre.vanrenssen@sydney.edu.au}{https://orcid.org/0000-0002-9294-9947}{This research was partially funded by the Australian Government through the Australian Research Council (project number DP240101353).}

\author{Shuei {Sakaguchi}}{The University of Sydney, Australia}{ssak8528@uni.sydney.edu.au}{https://orcid.org/0009-0008-3698-6090}{}

\authorrunning{A. van Renssen and S. Sakaguchi} 

\Copyright{André van Renssen and Shuei Sakaguchi} 

\ccsdesc[100]{{Theory of Computation $\rightarrow$ Computational Geometry}} 

\keywords{Ordered $\Theta$-graph, Local routing, Computational geometry} 

\category{} 

\relatedversion{} 

\acknowledgements{We thank the anonymous reviewers for their comments, which helped simplify the proofs.}

\nolinenumbers 

\EventEditors{John Q. Open and Joan R. Access}
\EventNoEds{2}
\EventLongTitle{42nd Conference on Very Important Topics (CVIT 2016)}
\EventShortTitle{CVIT 2016}
\EventAcronym{CVIT}
\EventYear{2016}
\EventDate{December 24--27, 2016}
\EventLocation{Little Whinging, United Kingdom}
\EventLogo{}
\SeriesVolume{42}
\ArticleNo{23}

\begin{document}

\maketitle

\begin{abstract}
    The problem of locally routing on geometric networks using limited memory is extensively studied in computational geometry. We consider one particular graph, the ordered $\Theta$-graph, which is significantly harder to route on than the $\Theta$-graph, for which a number of routing algorithms are known. Currently, no local routing algorithm is known for the ordered $\Theta$-graph.

    We prove that, unfortunately, there does not exist a deterministic memoryless local routing algorithm that works on the ordered $\Theta$-graph. This motivates us to consider allowing a small amount of memory, and we present a deterministic $O(1)$-memory local routing algorithm that successfully routes from the source to the destination on the ordered $\Theta$-graph. We show that our local routing algorithm converges to the destination in $O(n)$ hops, where $n$ is the number of vertices. To the best of our knowledge, our algorithm is the first deterministic local routing algorithm that is guaranteed to reach the destination on the ordered $\Theta$-graph.
\end{abstract}

\section{Introduction}

A \emph{geometric network} $G = (V, E)$ is a weighted graph where each vertex $v \in V$ is a point in the Euclidean plane $\mathbb{R}^2$ and each edge $(u, v) \in E$ is a straight line segment, connecting $u \in V$ and $v \in V$, weighted by the Euclidean distance $|uv|$. We define the \emph{weighted distance} $\delta_{G}(u,v)$ between two vertices $u$ and $v$ on $G$ to be the sum of the weights of the edges along the weighted shortest path from $u$ to $v$ in $G$.

A subgraph $H = (V, E')$ of a geometric network $G = (V, E)$ is called a $c$-spanner of $G$, if, for every pair of vertices $s,t \in V$, we have $\delta_{H}(s,t) \le c \cdot \delta_{G}(s,t)$, where $c \in \mathbb{R}_{\ge 1}$. Here, $G$ is called the \emph{underlying network} of $H$. The smallest constant $c$ such that $H$ is a $c$-spanner of $G$ is called the \emph{spanning ratio} or \emph{stretch factor} of $H$. We consider the situation where the underlying network $G$ is the \emph{complete Euclidean graph} of $V$ -- i.e., a $c$-spanner $H$ of $G$ approximates the Euclidean distance $|st| = \delta_{G}(s,t)$ between any pair of points $s,t \in V$ by a constant factor $c$. The spanning ratio of a class of spanners $\mathcal{G}$ is the spanning ratio of the worst-case instance $G \in \mathcal{G}$ maximizing the spanning ratio. See the textbook by Narasimhan and Smid~\cite{Narasimhan-Smid:2007:geometric-spanner-networks} and the survey by Bose and Smid~\cite{Bose-Smid:2013:survey} for a comprehensive overview of spanners and their open problems.

The study of geometric spanners is closely related to the design and analysis of efficient routing algorithms that forward a message between a pair of vertices. Given a source vertex $s$ and a target vertex $t$ on a graph $G$, a \emph{routing algorithm} aims to find a short path from $s$ to $t$. If the information of the entire graph can be known and kept track of during routing, several classical path-finding algorithms, including Dijkstra's algorithm~\cite{Dijkstra:1959:Graphs} and the Bellman-Ford-Moore algorithm~\cite{Moore:1959:ShortestPath}, can be used on spanners to find a short path. However, when we have restricted knowledge of the graph and limited memory, the routing problem becomes more challenging.

A routing algorithm is called \emph{$h$-local} if it only knows the $h$-neighborhood of the current vertex $u$ and the information of the destination $t$, where the $h$-neighborhood of $u$ is the set of vertices that can be reached from $u$ by taking at most $h$ distinct edges in the graph. A routing algorithm is called \emph{$m$-memory} if a memory of size $m$ is stored with the message of routing.\footnote{In this paper, we use the standard \emph{real RAM} model of computation used in computational geometry. In this model, each \emph{machine word} holds a real number or a $O(\log{n})$-bit integer (such as a pointer or an index), and we assume that arithmetic operations (excluding the floor and ceiling functions for real numbers) on machine words take $O(1)$ time. Consequently, a $O(1)$-memory routing algorithm may store $O(\log{n})$ bits.}

Two major classes of local routing algorithms have been actively researched in the literature: (1) deterministic $1$-local memoryless routing algorithms and (2) deterministic $1$-local $O(1)$-memory routing algorithms. Borrowing the notation used by Bose~et~al.~\cite{Bose-etal:2020:Theta6RoutingExpected}, these two classes of local routing algorithms can be formalised as follows:
\begin{itemize}
    \item A deterministic $1$-local memoryless routing algorithm on a graph $G = (V,E)$ is a \emph{routing function} $f : V \times V \times \mathcal{P}(V) \rightarrow V$, where $\mathcal{P}(\cdot)$ denotes the power set. We express the parameters of the function as $f(u, t, N(u))$, where $u \in V$ is the ``current'' vertex during routing (i.e., $u$ currently holds the message we want to route from the source to the destination), $t \in V$ is the destination vertex, and $N(u) \subseteq V$ is the $1$-neighborhood of $u$ (which simply means the set of immediate neighbors of $u$). The output of the routing function with the given parameters, $v = f(u, t, N(u)) \in N(u)$, is the neighbor of $u$ to which the message is forwarded.
    \item A deterministic $1$-local $O(1)$-memory routing algorithm on a graph $G = (V,E)$ is a routing function $g : V \times V \times \mathcal{P}(V) \times \mathcal{I} \rightarrow V$, which is similar to the routing function $f$ above, except the function $g(u,t,N(u),i)$ has an additional parameter $i \in \mathcal{I}$, modelling some information stored in the constant-sized memory of $g$ that can be used to make routing decisions.
\end{itemize}

A routing algorithm performed on a graph $H$ is called \emph{$c$-competitive} with respect to the underlying network $G$ if the total weight of the path traveled by the routing algorithm between any pair of vertices in $H$ is at most $c$ times the total weight of the shortest path in $G$~\cite{Bost-etal:2016:TightBoundsOnTheta}. The minimum possible value $c$ for which the given routing algorithm is $c$-competitive with respect to $G$ is called the \emph{routing ratio} of the routing algorithm in $H$.

\begin{figure}[htbp]
    \begin{center}
    \includegraphics[width=0.23\linewidth]{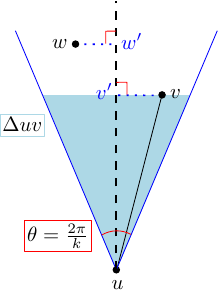}
    \caption{The creation of an edge $(u,v)$ during the construction of a $\Theta_{k}$-graph. The orthogonal projection $v'$ of $v$ is the closest to $u$ in the cone.}
    \label{fig:construct-Theta}
    \end{center}
\end{figure}

The $\Theta$-graph, introduced independently by Clarkson~\cite{Clarkson:1987:ThetaGraph} and Keil~\cite{Keil:1988:ThetaGraph}, is a geometric network studied extensively in the literature. Given a point set $V \subset \mathbb{R}^2$, the $\Theta_{k}$-graph $G = (V, E)$ is defined and constructed as follows: For each vertex $u \in V$, we let $k$ equally spaced rays emerge from $u$, resulting in $k$ cones, each with the aperture $\theta = \frac{2\pi}{k}$. For each cone of every vertex $u$, we add an edge from $u$ to the nearest neighbor $v$ within that cone, with distance measured along the bisector of the cone (see~\cref{fig:construct-Theta}).

It is known that $\Theta_{k}$-graphs are spanners (see \cref{tab:ThetaTable} for a summary). In addition, various local routing algorithms with good routing ratios have been developed for $\Theta_{k}$-graphs. A prime example is \emph{$\Theta$-routing}, also known as \emph{cone-routing}. $\Theta$-routing from $s$ to $t$ on a $\Theta_{k}$-graph is defined as follows: Let $u$ be the current vertex (initially, $u = s$). If there exists a direct edge from $u$ to $t$, follow this edge. Otherwise, follow the edge to the ``closest'' vertex in the cone of $u$ containing $t$. This is repeated until the destination $t$ is reached.

$\Theta$-routing is known to be competitive when used on $\Theta_{k}$-graphs for $k \ge 7$. For smaller $k$, other local routing algorithms have been developed (see~\cref{tab:ThetaTable} for a summary).

\begin{table}[htbp]
    \begin{center}
    \begin{tabular}{|c||c|c|}
        \hline
            & Spanning Ratio & Routing Ratio \\[1ex]
        \hline\hline
        $\Theta_{2}$ and $\Theta_{3}$ & Not spanners \cite{ElMolla:2009:YaoTheta} & Not spanners \cite{ElMolla:2009:YaoTheta} \\[1ex]
        \hline
        $\Theta_{4}$ & $17$ \cite{Bose-etal:2024:Theta4} & $17$ \cite{Bose-etal:2024:Theta4} \\[1ex]
        \hline
        $\Theta_{5}$ & $\frac{\sin{(3\pi/10)}}{\sin{(2\pi/5)}-\sin{(3\pi/10)}}$ \cite{Bose-etal:2024:Theta5} & - \\[1ex]
        \hline
        $\Theta_{6}$ & $2$ \cite{Bonichon-etal:2010:ThetaDelaunayConnection} & $2$ \cite{Bose-etal:2015:optimal-local-routing-Theta6} \\[1ex]
        \hline
        $\Theta_{(4i+2)}$ for $i \geq 2$ & $1+2\sin{(\frac{\theta}{2})}$ \cite{Bost-etal:2016:TightBoundsOnTheta} & $\frac{1}{1-2\sin{(\theta/2)}}$~\cite{Ruppert-Seidel:1991:ThetaGraph} \\[1ex]
        \hline
        $\Theta_{(4i+3)}$ for $i \geq 1$ & $\frac{\cos{(\theta/4)}}{\cos{(\theta/2)} - \sin{(3\theta/4)}}$ \cite{Bost-etal:2016:TightBoundsOnTheta} & $1+\frac{2\sin{(\theta/2)}\cdot\cos{(\theta/4)}}{\cos{(\theta/2)}-\sin{(\theta/2)}}$ \cite{Bost-etal:2016:TightBoundsOnTheta} \\[1ex]
        \hline
        $\Theta_{(4i+4)}$ for $i \geq 1$ & $1 + \frac{2\sin{(\theta/2)}}{\cos{(\theta/2)} - \sin{(\theta/2)}}$ \cite{Bost-etal:2016:TightBoundsOnTheta} & $1+\frac{2\sin{(\theta/2)}}{\cos{(\theta/2)}-\sin{(\theta/2)}}$ \cite{Bost-etal:2016:TightBoundsOnTheta} \\[1ex]
        \hline
        $\Theta_{(4i+5)}$ for $i \geq 1$ & $\frac{\cos{(\theta/4)}}{\cos{(\theta/2)} - \sin{(3\theta/4)}}$ \cite{Bost-etal:2016:TightBoundsOnTheta} & $1+\frac{2\sin{(\theta/2)}\cdot\cos{(\theta/4)}}{\cos{(\theta/2)}-\sin{(\theta/2)}}$ \cite{Bost-etal:2016:TightBoundsOnTheta} \\[1ex]
        \hline
    \end{tabular}
    \end{center}
    \caption{The current upper bounds on the spanning and routing ratio of $\Theta_{k}$-graphs, with $\theta = \frac{2\pi}{k}$.}
    \label{tab:ThetaTable}
\end{table}

Unfortunately, the $\Theta$-graph has a number of undesirable properties, including potentially a high maximum degree and a large hop-diameter. The \emph{ordered $\Theta$-graph} alleviated these issues. It was introduced by Bose et al.~\cite{Bose-etal:2004:OrderedThetaGraphs} in 2004 to achieve logarithmic maximum degree or logarithmic hop-diameter, in addition to the spanner property. An ordered $\Theta_{k}$-graph is constructed by inserting the vertices one by one and connecting each vertex to the ``closest'' \emph{previously} inserted vertex in each of the $k$ cones. The same paper showed that, for any vertex set, inserting vertices in a random order produces with high probability an ordered $\Theta$-graph with logarithmic hop-diameter, and constructed a specific insertion order that deterministically produces an ordered $\Theta$-graph with logarithmic maximum degree.

However, achieving these additional properties comes at a price. For example, ordered $\Theta$-graphs with $4$, $5$, or $6$ cones are not spanners~\cite{Bose-etal:2016:ThePriceOfOrder}, despite the fact that their unordered counterparts are. In addition, for larger cone numbers, ordered $\Theta$-graphs have worse spanning ratios compared to $\Theta$-graphs. The best known upper bounds and lower bounds on the spanning ratio of ordered $\Theta_{k}$-graphs are summarized in~\cref{tab:OrderedThetaTable}.

\begin{table}[htbp]
    \begin{center}
    \begin{tabular}{|c||c|c|}
        \hline
            & Upper Bound & Lower Bound \\[1ex]
        \hline\hline
        $\Theta_{3}$, $\Theta_{4}$, $\Theta_{5}$, and $\Theta_{6}$ & Not spanners \cite{Bose-etal:2016:ThePriceOfOrder} & Not spanners \cite{Bose-etal:2016:ThePriceOfOrder} \\[1ex]
        \hline
        $\Theta_{(4i+2)}$ for $i \geq 2$ & $\frac{1}{1-2\sin{(\theta/2)}}$ \cite{Bose-etal:2004:OrderedThetaGraphs} & $\frac{1}{1-2\sin{(\theta/2)}}$ \cite{Bose-etal:2016:ThePriceOfOrder} \\[1ex]
        \hline
        $\Theta_{(4i+3)}$ for $i \geq 1$ & $\frac{1}{1-2\sin{(\theta/2)}}$ \cite{Bose-etal:2004:OrderedThetaGraphs} & $\frac{\cos{(\theta/4)}+\sin{(\theta)}}{\cos{(3\theta/4)}}$ \cite{Bose-etal:2016:ThePriceOfOrder} \\[1ex]
        \hline
        $\Theta_{(4i+4)}$ for $i \geq 1$ & $1+\frac{2\sin{(\theta/2)}}{\cos{(\theta/2)}-\sin{(\theta/2)}}$ \cite{Bose-etal:2016:ThePriceOfOrder} & $1+\frac{2\sin{(\theta/2)}}{\cos{(\theta/2)}-\sin{(\theta/2)}}$ \cite{Bose-etal:2016:ThePriceOfOrder} \\[1ex]
        \hline
        $\Theta_{(4i+5)}$ for $i \geq 1$ & $\frac{1}{1-2\sin{(\theta/2)}}$ \cite{Bose-etal:2004:OrderedThetaGraphs} & $1+\frac{2\sin{(\theta/2)}\cdot\cos{(\theta/4)}}{\cos{(\theta/2)}-\sin{(3\theta/4)}}$ \cite{Bose-etal:2016:ThePriceOfOrder} \\[1ex]
        \hline
    \end{tabular}
    \end{center}
    \caption{The upper and lower bounds on the spanning ratio of ordered $\Theta_{k}$-graphs, with $\theta = \frac{2\pi}{k}$.}
    \label{tab:OrderedThetaTable}
\end{table}

\begin{figure}[htbp]
    \begin{center}
    \includegraphics[width=0.35\linewidth]{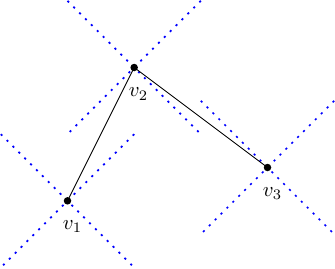}
    \caption{Three vertices $v_1$, $v_2$, and $v_3$ are inserted in the order $v_1, v_2, v_3$. When inserting $v_2$, an edge to $v_1$ is added. When inserting $v_3$, only an edge to $v_2$ is added, resulting in $v_1$ having vertex $v_3$ in a cone with no edges.}
    \label{fig:theta-routing-fails}
    \end{center}
\end{figure}

Although we have good local routing algorithms for the $\Theta$-graph, no local routing algorithm is known for the ordered $\Theta$-graph. A major obstacle is that even if there are vertices in a cone, there may still not be any edges, due to the insertion order (see~\cref{fig:theta-routing-fails}). This makes the ordered $\Theta$-graph much more challenging to route on than the $\Theta$-graph -- for example, $\Theta$-routing fails on the ordered $\Theta$-graph.

\section{Technical overview}

Our contribution is twofold: First, we show why so far no deterministic \emph{memoryless} local routing algorithms have been found. Specifically, we prove that there does not exist a deterministic memoryless $h$-local routing algorithm on the ordered $\Theta_{k}$-graph, for $h \geq 1$ and $k \geq 2$. We construct two ordered $\Theta_{k}$-graphs $L_{h}^{k}$ and $R_{h}^{k}$ that cannot be distinguished in this routing model, and show that any such algorithm fails on one of $L_{h}^{k}$ and $R_{h}^{k}$. By doing so, we obtain the following theorem.

\begin{restatable}{theorem}{thmImpossibilityResult} \label{thrm:impossibility-result}
    There is no deterministic memoryless $h$-local routing algorithm (for any integer $h \geq 1$) capable of finding a path between every pair of source vertex $s$ and destination vertex $t$ on the ordered $\Theta_{k}$-graph (for any integer $k \geq 2$).
\end{restatable}

We then present a deterministic $O(1)$-memory $1$-local routing algorithm $\mathcal{A}$ and prove that it converges to the destination in $O(n)$ hops (i.e., taking at most $O(n)$ edges), where $n = |V|$. Our algorithm $\mathcal{A}(s, t)$ works in two phases: The first routes from the source $s$ to the first inserted vertex $v_1$ by repeatedly moving to a neighbor with a smaller order. In the second phase, our algorithm routes from $v_1$ to the destination $t$ by systematically exploring vertices that might lead to $t$, while ensuring that there is no need to explicitly keep track of the exploration path. We use a simple memoryless $1$-local routing algorithm we call \emph{ordered $\Theta$-routing}, which allows us to implicitly keep track of the vertices we need to backtrack to by utilizing the geometric property of the ordered $\Theta$-graph.

To prove that our algorithm $\mathcal{A}(s, t)$ converges, we consider the ordered $\Theta$-routing path from $t$ to $v_1$ and show that our algorithm explores this path in the reverse direction, by arguing that the second phase of our algorithm follows the same edges as a depth-first search on a spanning tree of the ordered $\Theta$-graph, such that the tree is the union of all ordered $\Theta$-routing paths towards $v_1$. Then, we conclude that our algorithm reaches the destination in at most $O(n)$ hops and $O(kn)$ time, obtaining the following theorem.

\begin{restatable}{theorem}{thmAlgorithmA} \label{thrm:algorithm-A}
    Given an arbitrary source vertex $s \in V$ and the destination vertex $t \in V$ of an ordered $\Theta_{k}$-graph $G = (V,E)$ (for an integer $k \geq 2$), the algorithm $\mathcal{A}(s, t)$ is a deterministic $O(1)$-memory $1$-local routing algorithm which successfully routes from $s$ to $t$ in $O(n)$ hops. In particular, $\mathcal{A}(s, t)$ takes $O(kn)$ time in the worst-case.
\end{restatable}

\section{Preliminaries}

\begin{figure}[htbp]
    \begin{center}
    \includegraphics[width=0.9\linewidth]{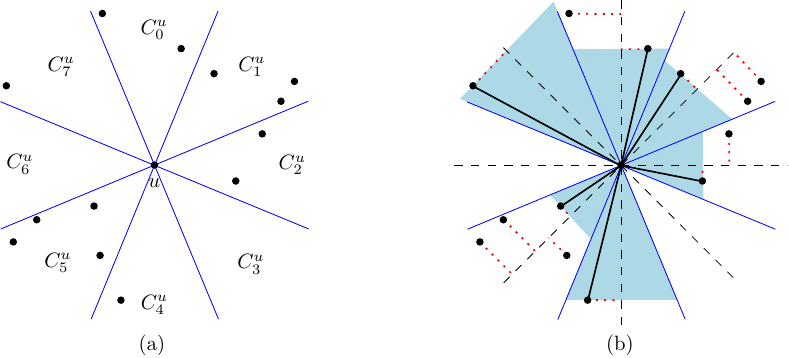}
    \caption{(a) The numbered cones around $u$ and (b) the edges constructed.}
    \label{fig:ordered-theta-cones-edges}
    \end{center}
\end{figure}

Given a set of vertices $V \subset \mathbb{R}^2$ in the Euclidean plane, the ordered $\Theta_{k}$-graph $G = (V, E)$ is a geometric network, with edges being undirected straight line segments, constructed as follows. For each vertex $u \in V$, we define $\rho(u)$ to be the order of insertion of $u$ -- i.e., if $v_i$ is the $i$-th vertex inserted into $G$ for some positive integer $i \leq |V|$, we let $\rho(v_i) = i$. We incrementally insert the vertices from the first to the last according to the order $\rho$. Upon inserting each vertex $u \in V$, we partition the plane around $u$ into $k$ disjoint regions, by projecting $k$ equally spaced rays from $u$, such that the angle between each pair of consecutive rays is $\theta = \frac{2\pi}{k}$ (see~\cref{fig:ordered-theta-cones-edges}(a)). We define a \emph{cone} $C^{u}$ of $u$ to be a region in the plane between two consecutive rays originating from $u$. We orient the $k$ cones such that the bisector of one of them coincides with the vertical half-line emerging from $u$. Let the cone with such a bisector be $C_{0}^{u}$, from which we start numbering the cones in clockwise order around $u$. For the insertion of every other vertex in $V$, we orient and number the cones around the vertex in the same way. We write $C_{i}^{u}$ to denote the $i$-th cone of $u$, for $0 \leq i < k$. We say that $u$ is the \emph{apex} of $C_{i}^{u}$. We also define $C^{u}(v)$ to be the cone of $u$ containing vertex $v \ne u$.

Upon the insertion of each vertex $u \in V$, an edge is created to the ``closest'' \emph{previously} inserted vertex in each cone $C_{i}^{u}$ of $u$ (see~\cref{fig:ordered-theta-cones-edges}(b)). More formally, an edge is created from $u$ to $v$ if (1) $u$ is the apex of the cone that contains $v$ such that $\rho(v) < \rho(u)$ and (2) for every vertex $w$ within the cone such that $\rho(w) < \rho(u)$, we have $|uv'| \leq |uw'|$, where $v'$ (resp. $w'$) denotes the orthogonal projection of $v$ (resp. $w$) onto the bisector of the cone. The canonical triangle $\Delta{uv}$ corresponding to the edge created from $u$ to $v$ is defined by the boundary of cone $C^{u}(v)$ and the line through $v$ and $v'$ (see the regions in light blue in~\cref{fig:ordered-theta-cones-edges}(b)). Note that only the vertices inserted before $u$ are considered for the edge creation process, and thus different orderings of vertex insertions result in different ordered $\Theta_{k}$-graphs even if the vertex set is the same.

For the insertion of each vertex $u \in V$, we store its order $\rho(u)$ on the vertex, in addition to the $x$- and $y$-coordinates of $u$. Since $1 \leq \rho(u) \leq n$ (where $n = |V|$), storing the order $\rho(u)$ requires at most $\lceil \log{n} \rceil$ extra bits per vertex, fitting in a single machine word.

It is important to note that a canonical triangle of the ordered $\Theta$-graph is not necessarily empty, despite the fact that any canonical triangle corresponding to an edge in the $\Theta$-graph is always empty of vertices in its interior, since the vertices are inserted incrementally in the ordered $\Theta$-graph. Suppose, during the construction of an ordered $\Theta$-graph, we have just inserted a vertex $u \in V$ with order $\rho(u)$. Let $v$ with order $\rho(v) < \rho(u)$ be the closest neighbor to $u$ in cone $C^{u}(v)$. At this stage, canonical triangle $\Delta{uv}$ is empty. However, later we can insert a vertex $w$ with order $\rho(w) > \rho(u)$ into the interior of $\Delta{uv}$, making the canonical triangle non-empty.

For simplicity, we assume that $V$ is in \emph{general position}. Specifically, we will assume that, upon inserting each vertex during the construction of the ordered $\Theta$-graph, it is clear which vertices lie in which cone (i.e., no previously inserted vertex lies on a boundary of a cone of the currently inserted vertex) and it is clear which vertex is the ``closest'' in each cone of a vertex (i.e., no two vertices in the cone share the same orthogonal projection). This assumption can be removed using standard techniques.

\subsection{Connectedness}
Before considering the routing question on the ordered $\Theta$-graph, we consider its connectedness. When $k > 6$, the ordered $\Theta_{k}$-graph is a spanner~\cite{Bose-etal:2016:ThePriceOfOrder}, which implies connectedness. When $k \in \{4,5,6\}$, the ordered $\Theta_{k}$-graph is not a spanner, and no formal proof that it is connected seems to be known. So, as a warm-up, we will first give a simple inductive proof for the connectedness of the ordered $\Theta_{k}$-graph for any $k \geq 2$.

\begin{theorem}[Connectedness]\label{thrm:connectedness}
    The ordered $\Theta_{k}$-graph $G = (V,E)$ for $k \geq 2$ with $n \geq 1$ vertices is connected.
\end{theorem}

\begin{proof}
    We prove this by induction on the number of vertices in the graph. The base case is when $G$ has only a single vertex $(n = 1)$ -- trivially, any graph with a single vertex is connected. The inductive case is when there are $n > 1$ vertices. As our induction hypothesis, assume that $G$ is connected when it has $i \geq 1$ vertices $V = \{v_1, \dots, v_i\}$. Consider what happens if we insert a new $(i+1)$-th vertex $v_{i+1}$ into $G$. Since all of the previously inserted $i$ vertices lie somewhere in the plane, each of them exists in some cone of the newly inserted vertex $v_{i+1}$. So, upon the insertion of $v_{i+1}$ into $G$, an edge is created from $v_{i+1}$ to some vertex $v \in \{v_1, \dots, v_i\}$. Since $G$ with the vertex set $\{v_1, \dots, v_i\}$ is connected (by the induction hypothesis) and the newly inserted vertex $v_{i+1}$ is linked to a vertex $v \in \{v_1, \dots, v_i\}$, the new ordered $\Theta_{k}$-graph $G' = (V',E')$, where $V' = \{v_1, \dots, v_i, v_{i+1}\}$, is also connected.
\end{proof}

\section{Local Routing on Ordered $\Theta_{k}$-graphs}

In this section, we will show that no deterministic \emph{memoryless} local routing algorithm is guaranteed to reach the destination on the ordered $\Theta_{k}$-graph, for any integer $k \geq 2$ (\cref{subsec:0-memory}). As this motivates us to allow a small amount of memory, we will then present a deterministic \emph{$O(1)$-memory} local routing algorithm that is guaranteed to reach the destination (\cref{subsec:constant-memory}).

We first define an important operation used by local routing algorithms.

\begin{definition}[Hop] \label{def:hop}
    We define a \emph{hop} to be a single move or a forwarding operation from one vertex to its neighbor. Namely, taking a hop means we either (1) move from a vertex $u$ to another vertex $v$ through an edge, or (2) forward a message from $u$ to $v$ via an edge between them.
\end{definition}

\subsection{Impossibility result for memoryless local routing} \label{subsec:0-memory}

We show that no deterministic memoryless local routing algorithm works on the ordered $\Theta$-graph, even if the algorithm is $h$-local, and thus can access the $h$-neighborhood, for $h \geq 1$. This impossibility result naturally motivates us to allow a small amount of memory.

\thmImpossibilityResult*

\begin{proof}
    We present two ordered $\Theta_{k}$-graphs such that any deterministic memoryless $h$-local routing algorithm $A$ fails to route from $s$ to $t$ on one of these graphs. Each of these two graphs has $2h+3$ vertices and $2h+2$ edges, forming a simple path (see~\cref{fig:LR(h-local)ThetaK}).

    \begin{figure}[htbp]
        \begin{center}
        \includegraphics[width=0.75\linewidth]{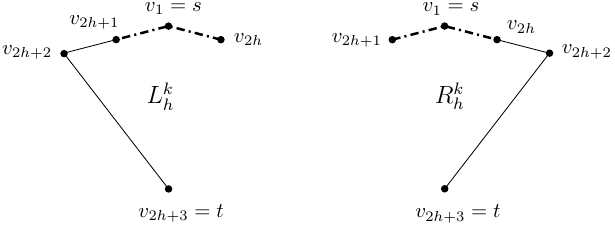}
        \caption{A pair of ordered $\Theta_{k}$-graphs where any deterministic memoryless $h$-local routing algorithm fails on one of them.}
        \label{fig:LR(h-local)ThetaK}
        \end{center}
    \end{figure}

    Intuitively, we design two graphs, where one has a path from $s$ to $t$ to the left of $s$ and $t$ ($L_{h}^{k}$) and one to the right of $s$ and $t$ ($R_{h}^{k}$). By ensuring that these graphs are identical up to $h$ hops from $s$, no $h$-local routing algorithm can tell them apart, and since $A$ is deterministic and has no memory, it will always make the same choice at $s$, regardless of what path of the graph it has explored in the past.

    Formally, we specify the location of the $2h+3$ vertices of $L_{h}^{k}$ as follows: $v_{1} = (0,0)$, $v_{2i} = (\frac{i}{h}, -i\epsilon)$ and $v_{2i+1} = (-\frac{i}{h}, -i\epsilon)$ for $i \in \{1,2,\dots,h\}$, $v_{2h+2} = (-2, -2h\epsilon)$, and $v_{2h+3} = (0, -\frac{2}{\tan{(\theta/2)}} -3h\epsilon)$. Note that the $y$-coordinate of $v_{2h+3}$ is given in terms of $\theta$, $\epsilon$, and $h$ such that all previously inserted vertices belong to $C_{0}^{v_{2h+3}}$. Here, $\epsilon$ is an arbitrarily small non-negative real number, such that for $j \in \{2,3,\dots,2h+3\}$, a cone of $v_{j}$ contains all vertices $v_{j'}$ for $j' \in \{1,2\dots,j-1\}$. In particular, if no boundary of the cones of a vertex $u$ coincides with the horizontal line through $u$, then $\epsilon$ can simply be $0$. We insert these vertices in ascending order: $v_{1}$ and then $v_{j}$ for $j \in \{2, \dots, 2h+3\}$. The creation of $R_{h}^{k}$ is very similar to that of $L_{h}^{k}$. The only difference is that $v_{2h+2}$ is placed at $(2, -2h\epsilon)$ in $R_{h}^{k}$.

    We set $s$ to be $v_{1}$ and $t$ to be $v_{2h+3}$ for both $L_{h}^{k}$ and $R_{h}^{k}$, then we run $A$ on both graphs. Consider the current vertex $u = s = v_{1}$ of routing. From the perspective of $A$, the local configuration at $v_{1}$ in $L_{h}^{k}$ and the local configuration at $v_{1}$ in $R_{h}^{k}$ are indistinguishable: the coordinates and ordering information of $u = v_{1}$, $t = v_{2h+3}$, and the $h$-neighborhood of $u = v_{1}$ are identical for both graphs. Hence, any deterministic memoryless $h$-local routing algorithm $A$ must make the same choice at $v_{1}$ in both $L_{h}^{k}$ and $R_{h}^{k}$. If $A$ decides to forward the message to $v_{2}$ from $v_{1}$, $A$ can never reach $v_{3}$ and therefore $t$ in $L_{h}^{k}$, since $A$ always deterministically makes the same choice at $v_{1}$ to forward the message to $v_{2}$ because the algorithm is memoryless. Similarly, if $A$ decides to forward the message to $v_{3}$ from $s = v_{1}$, $A$ cannot reach $v_2$ and thus $t$ on $R_{h}^{k}$.
\end{proof}

\subsection{$O(1)$-memory local routing} \label{subsec:constant-memory}

Now that we know we cannot route locally without using some memory, we shift to having a limited amount of it. In particular, our approach will store $O(1)$ information.

Our algorithm works in two phases: first we route to the lowest order vertex $v_1$ (\cref{subsubsec:first-part}), and then to $t$ (\cref{subsubsec:second-part}). The first phase repeatedly moves from the current vertex to a neighbor with a smaller order, which results in reaching $v_1$ starting from $s$ without knowing the location of $v_1$ in advance. The second phase routes from $v_1$ to $t$ by systematically exploring vertices that might lead to $t$ while implicitly keeping track of the vertices we need to backtrack to by utilizing the geometric properties of the ordered $\Theta$-graph.

Before describing our algorithm, we first present its core component, which we call \emph{ordered $\Theta$-routing}, and prove a geometric property of the ordered $\Theta$-graph in relation to ordered $\Theta$-routing. This simple algorithm relates to our main algorithm in two ways:
\begin{enumerate}
    \item The geometric property of ordered $\Theta$-routing relating to the graph ensures the correctness of the first phase of our main algorithm. However, note that we do not run ordered $\Theta$-routing during the first phase, as we do not know the location of $v_1$.
    \item Ordered $\Theta$-routing will be used as the core subroutine of the second phase of our main algorithm, particularly for backtracking purposes. This allows us to implicitly keep track of the ``exploration path'' from $v_1$ to the current vertex $u$ without storing the path.
\end{enumerate}

Given an ordered $\Theta$-graph $G$ and an arbitrary pair of source and destination vertices $s,t \in V$, we define ordered $\Theta$-routing as follows.

\begin{algorithm}[Ordered $\Theta$-routing]
    Initialize the current vertex $u \leftarrow s$. Then, we repeat this forwarding operation until reaching $t$ or ``getting stuck'': If there exists an edge from $u$ to $t$, take that edge -- we are done. Otherwise, take the edge to the closest neighbor $w \in \Delta{ut}$ such that $\rho(u) > \rho(w)$. We call this single hop an \emph{ordered $\Theta$-routing step}. If both conditions are negative -- i.e., (1) there does not exist an edge from $u$ to $t$ and (2) there does not exist an edge from $u$ to a vertex $w \in \Delta{ut}$ with a smaller order -- ordered $\Theta$-routing is considered ``stuck'' and we terminate.
\end{algorithm}

The ordered $\Theta$-routing scheme we presented is inspired by the $\Theta$-routing scheme developed for the regular $\Theta$-graph. The main difference is that ordered $\Theta$-routing considers the ordering information when determining which neighbor to forward the message to -- it always chooses the closest neighbor with a smaller order in the cone containing the destination.

Note that ordered $\Theta$-routing from the source $s$ to the destination $t$ can only forward the message at the current vertex $u$ to the next vertex if either (1) $u$ has $t$ as a neighbor or (2) $u$ has a neighbor $w \ne t$ lying in $\Delta{ut}$ such that $\rho(u) > \rho(w)$. Thus, it can fail to reach the destination when neither condition is met. However, as we will see below, ordered $\Theta$-routing is guaranteed to reach the destination if the destination is $v_1$, the first inserted vertex during the construction of the ordered $\Theta$-graph.

\begin{lemma} \label{lmm:ordered-theta-routing-towards-v1}
    Given an arbitrary source $s \in V$ and the first inserted vertex $v_1 \in V$ of any ordered $\Theta_{k}$-graph $G = (V,E)$ ($k \geq 2$), ordered $\Theta$-routing towards $v_1$ always reaches $v_1$ through a simple path, which we call the \emph{ordered $\Theta$-routing path} from $s$ to $v_1$.
\end{lemma}

\begin{proof}
    We prove the lemma by routing from $s$ to $v_1$ using ordered $\Theta$-routing (see~\cref{fig:ordered-theta-routing-towards-v1}). We initialize the current vertex $u \leftarrow s$, and consider the two possible routing steps:
    \begin{itemize}
        \item If there is an edge from $u$ to $v_1$, we can take that edge and reach $v_1$ -- we are done.
        \item Otherwise, consider the canonical triangle $\Delta{uv_1}$. The fact that there does not exist an edge from $u$ to $v_1$ implies that, when the current vertex $u$ was inserted during the construction of the graph, there existed a vertex $w \in \Delta{uv_1}$ inserted before $u$ (i.e., we have $\rho(u) > \rho(w)$), such that $w$ was the closest neighbor in cone $C^{u}(v_1)$, implying the construction of edge $\{u, w\}$, which can be taken by ordered $\Theta$-routing.
    \end{itemize}

    \begin{figure}[htbp]
        \begin{center}
        \includegraphics[width=0.5\linewidth]{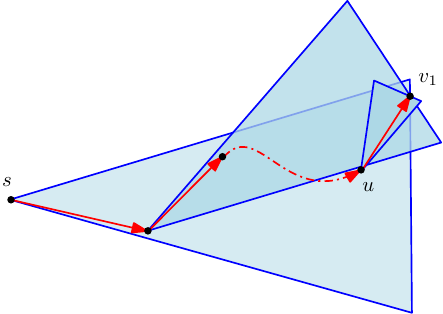}
        \caption{The ordered $\Theta$-routing path from $s$ to $v_1$, which is a simple path.}
        \label{fig:ordered-theta-routing-towards-v1}
        \end{center}
    \end{figure}

    Ordered $\Theta$-routing from $s$ to $v_1$ eventually converges. Since the order $\rho(v_1) = 1$ is the smallest among all vertices in $V$, the current vertex $u$ has an edge either to $v_1$ or to a vertex $w \in \Delta{uv_1}$ such that $\rho(u) > \rho(w)$ by construction. So, the order $\rho(u)$ of the current vertex $u$ keeps decreasing by at least $1$ for each step. By monotonicity, within $n-1$ hops, the canonical triangle $\Delta{uv_1}$ becomes empty of vertices whose order is smaller than that of $u$, which implies there exists an edge from $u$ to $v_1$. In particular, the ordered $\Theta$-routing path from $s$ to $v_1$ is simple, as we go through vertices in monotonically decreasing order.
\end{proof}

\subsubsection{Reaching vertex $v_1$} \label{subsubsec:first-part}

Note that we can route from $s$ to $v_1$ using ordered $\Theta$-routing if $v_1$ is explicitly given to us or every vertex stores $v_1$. However, we are only given $s$ and $t$ and we want to minimize space, so we present another simple algorithm, $\mathcal{A}^{down}$, that routes from $s$ to $v_1$ without knowing $v_1$ in advance, by utilizing the property that every vertex has a neighbor with a smaller order number by~\cref{lmm:ordered-theta-routing-towards-v1}.

\begin{algorithm}[$\mathcal{A}^{down}(s)$] \label{algo:A-down}
    Given a source $s$, we initialize the current vertex $u$ to be $s$. Then, we repeat the following process until we have $u = v_1$: Pick a neighbor $w \in N(u)$ such that $\rho(u) > \rho(w)$ (if there are multiple options, pick an arbitrarily one among them, say the minimum) and update $u \leftarrow w$.
\end{algorithm}

Below, we analyze the correctness and complexity of $\mathcal{A}^{down}$. Let $n = |V|$.

\begin{lemma} \label{lmm:routing-to-the-first-inserted-vertex}
    Given an arbitrary source vertex $s \in V$ of any ordered $\Theta_{k}$-graph $G = (V,E)$ ($k \geq 2$), $\mathcal{A}^{down}(s)$ is a deterministic memoryless $1$-local routing algorithm which always reaches the first inserted vertex $v_1 \in V$ in $O(n)$ hops. In particular, $\mathcal{A}^{down}(s)$ takes $O(kn)$ time in the worst-case.
\end{lemma}

\begin{proof}
    We first show that $\mathcal{A}^{down}$ always converges to $v_1$.
    Consider the current vertex $u \ne v_1$ (initially $u = s$) while executing $\mathcal{A}^{down}$. By~\cref{lmm:ordered-theta-routing-towards-v1}, $u$ always has a neighbor $w$ such that $\rho(u) > \rho(w)$, unless $u = v_1$ (see~\cref{fig:routing-towards-v1}). As a result, $\mathcal{A}^{down}$ can keep updating $u$ to be a neighbor with a smaller order. Since the order of a vertex cannot go below $\rho(v_1) = 1$, the process ends when $u = v_1$.

    \begin{figure}[htbp]
        \begin{center}
        \includegraphics[width=0.53\linewidth]{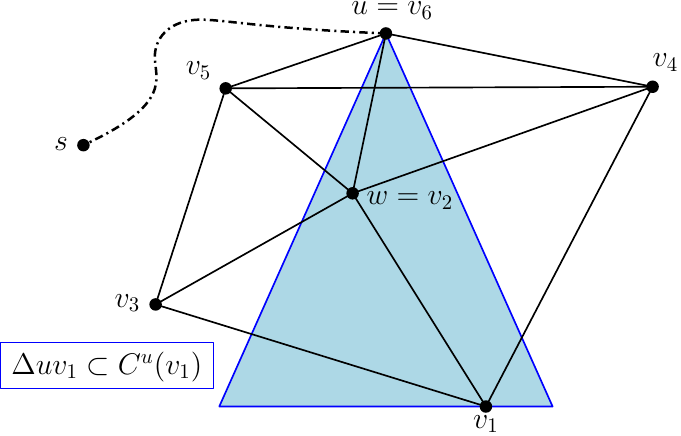}
        \caption{$u$ always has a neighbor $w$ such that $\rho(u) > \rho(w)$. For simplicity, $\rho(v_i) = i$ for $i \in [n]$.}
        \label{fig:routing-towards-v1}
        \end{center}
    \end{figure}

    To bound the number of hops, we observe that there are $n = |V|$ vertices in $G$ and each hop decreases the order $\rho(u)$ by at least $1$, so there can be at most $n - 1$ hops.

    It remains to analyze the total running time. At any current vertex $u$, $\mathcal{A}^{down}$ checks the neighbors $N(u)$ of $u$ until finding one with a smaller order. Once $\mathcal{A}^{down}$ takes a hop from $u$ to such a neighbor, $u$ is never visited again, because the order of the current vertex monotonically decreases, preventing a cycle. So, the overall number of operations for checking the neighbors of every current vertex visited throughout the execution of $\mathcal{A}^{down}$ is upper bounded by $\sum_{u \in V} \text{deg(u)} = 2|E|$. Since the construction of the ordered $\Theta_{k}$-graph $G = (V,E)$ ensures that $|E| < kn$, the total running time is $O(kn)$.

    Finally, we observe that $\mathcal{A}^{down}$ is deterministic, memoryless, and $1$-local by construction.
\end{proof}

Now that we can route from $s$ to $v_1$ using $\mathcal{A}^{down}(s)$, we move on to showing how to route from $v_1$ to $t$.

\subsubsection{Finding the destination $t$} \label{subsubsec:second-part}

Before describing how to route from $v_1$ to $t$, we consider the union of all ordered $\Theta$-routing paths to $v_1$ in the ordered $\Theta$-graph $G = (V,E)$, which will inform our routing strategy.

Let $P_u$ be the set consisting of the edges of the ordered $\Theta$-routing path from vertex $u$ to $v_1$, and let $Q$ be the set of edges of all ordered $\Theta$-routing paths to $v_1$, i.e., $Q = \bigcup_{u \in V} P_u$.
Let $T = G[Q]$ be the subgraph of $G$ induced by the edge set $Q$. Observe that by definition the vertex set of $T$ equals $V$. By~\cref{lmm:ordered-theta-routing-towards-v1}, $T$ is a connected spanning subgraph of $G$. We now argue that $T$ is a tree.

\begin{lemma} \label{lmm:spanning-tree}
    $T$ is a spanning tree of $G$.
\end{lemma}

\begin{proof}
    Since we already know that $T$ is a spanning subgraph of $G$, we will show that $T$ has $|V| - 1$ edges in total, to conclude that $T$ is a spanning tree of $G$. To do so, we iterate through all edges of $T$, charging each edge of $T$ to the endpoint with a greater order.

    Observe that $v_1$ is not charged any edge since $\rho(v_1) = 1$ is the smallest order of all vertices.
    We argue that every vertex $u \ne v_1$ is charged exactly one edge.
    By the argument of~\cref{lmm:ordered-theta-routing-towards-v1}, an ordered $\Theta$-routing step at any vertex $u \ne v_1$ towards $v_1$ leads to a neighbor $w \in N(u)$ such that $\rho(u) > \rho(w)$ (see~\cref{fig:tree}). Thus, edge $e = \{u, w\}$ is charged to $u$.

    \begin{figure}[htbp]
        \begin{center}
        \includegraphics[width=0.45\linewidth]{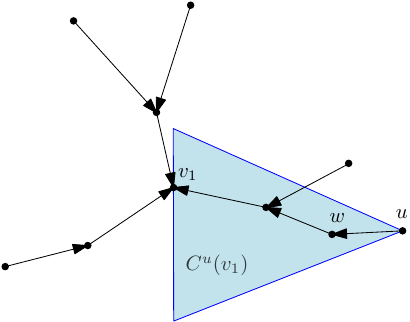}
        \caption{An ordered $\Theta$-routing step at $u$ towards $v_1$ leads to a single neighbor $w \in C^{u}(v_1)$.}
        \label{fig:tree}
        \end{center}
    \end{figure}

    We argue that no edge other than $e$ is charged to $u$. For the sake of contradiction, suppose $T$ has another edge $e' = \{u, w'\}$ such that $w' \ne w$ and $\rho(u) > \rho(w')$, so $e'$ will also be charged to $u$. Then, by the definition of $T$, an ordered $\Theta$-routing step at $u$ towards $v_1$ leads to both $w$ and $w'$, which contradicts the definition of ordered $\Theta$-routing.

    Consequently, as we iterate through all edges of $T$, every vertex $u \in V$ except for $v_1$ is charged exactly one edge of $T$, so $T$ has $|V|-1$ edges in total, which implies $T$ is a spanning tree of $G$.
\end{proof}

We now present $\mathcal{A}^{up}$, a $1$-local algorithm that routes from $v_1$ to $t$. Intuitively, the algorithm finds the reverse of the ordered $\Theta$-routing path from $t$ to $v_1$ by performing exploration and backtracking operations in the same way as a depth-first search on $T$. In other words, the algorithm explores all ordered $\Theta$-routing paths towards $v_1$ ``backwards'' until it finds $t$.

\begin{algorithm}[$\mathcal{A}^{up}(v_1, t)$]
    ~
    \begin{enumerate}
        \item Initialize the previous vertex $u^p$ to be $null$.
        \item Initialize the current vertex $u$ to be $v_1$.
        \item Initialize a bit named $state$ to be $1$, indicating whether the previous hop was exploration or backtracking.
        \item Repeat this block until $u = t$:
        \begin{itemize}
            \item If $state = 1$ (the previous hop was exploration):\\
                Let $l$ be the horizontal half-line originating from $u$ towards the right. Sweep $l$ in counter-clockwise direction (see~\cref{fig:counter-clockwise-exploration}). Let $v \in N(u)$ be the first neighbor that touches $l$. Check the following two conditions:
                \begin{enumerate}
                    \item $\rho(v) > \rho(u)$.
                    \item $u \in C^{v}(v_1)$.
                \end{enumerate}
                If both conditions are met, then we explore $v$ from $u$ by updating the current vertex $u$, i.e., we set $u \leftarrow v$. Otherwise, we keep sweeping $l$ in counter-clockwise direction, looking for a neighbor of $u$ meeting both conditions -- if one is found, explore that neighbor, updating $u$. If $l$ reaches its original position, we perform an ordered $\Theta$-routing step at $u$ towards $v_1$ to backtrack to a vertex $w$ (observe that we previously explored $u$ from $w$), updating the previous vertex $u^p \leftarrow u$, the current vertex $u \leftarrow w$, and the state bit $state \leftarrow 0$.
            \item Else, i.e., $state = 0$ (the previous hop was backtracking):\\
                Let $l$ be the half-line originating from $u$ through $u^p$. Sweep $l$ in counter-clockwise direction as usual, starting from this position, until it becomes the half-line pointing to the right. For each $v \in N(u)$ that touches $l$, check whether it meets the same two conditions mentioned in the previous case. If such a neighbor is found, explore it and update the variables accordingly, setting $u$ to be the neighbor and $state$ to be $1$. If $l$ becomes the half-line pointing to the right, we perform an ordered $\Theta$-routing step at $u$ towards $v_1$ to backtrack to a vertex $w$, updating the variables as in the previous case, setting $u^p \leftarrow u$, $u \leftarrow w$, and $state \leftarrow 0$.
        \end{itemize}
    \end{enumerate}
\end{algorithm}

\begin{figure}[htbp]
    \begin{center}
    \includegraphics[width=0.5\linewidth]{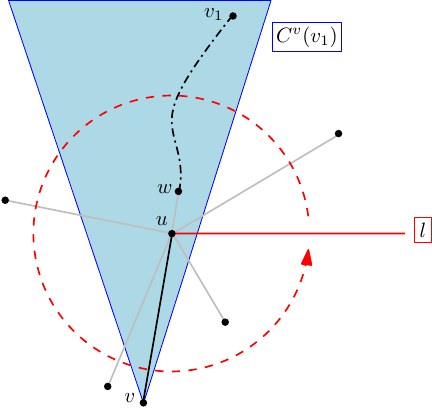}
    \caption{The counter-clockwise exploration of $\mathcal{A}^{up}(v_1, t)$.}
    \label{fig:counter-clockwise-exploration}
    \end{center}
\end{figure}

We first justify that $\mathcal{A}^{up}$ is a deterministic $O(1)$-memory $1$-local routing algorithm: There is no randomness involved in the algorithm, so it is deterministic. Also, the algorithm only stores $v_1$, $u^p$, and the $state$ bit, so it only uses $O(1)$-memory. Finally, the algorithm only accesses the $1$-neighborhood $N(u)$ of the current vertex $u$, so it is $1$-local. Therefore, we obtain the following lemma.

\begin{lemma} \label{lmm:deterministic-constant-memory-1-local}
    $\mathcal{A}^{up}$ is a deterministic $O(1)$-memory $1$-local routing algorithm.
\end{lemma}

We now argue that $\mathcal{A}^{up}(v_1, t)$ is essentially a depth-first search on $T$.
To do so, we show that the algorithm only takes edges of $T$.
By definition any backtracking step takes an edge of $T$.
So, we will show that each exploration step takes an edge of $T$.
In particular, we show that it takes an ordered $\Theta$-routing edge backwards.
To do so, we prove that the algorithm explores a neighbor $v$ of the current vertex $u$ (i.e., we have $\rho(v) > \rho(u)$ and $u \in C^{v}(v_1)$) if and only if an ordered $\Theta$-routing step at $v$ towards $v_1$ leads to $u$.

\begin{lemma} \label{lmm:two-conditions}
    Let $v \in V$ be a neighbor of $u \in V$. An ordered $\Theta$-routing step at $v$ towards $v_1$ leads to $u$ if and only if $\rho(v) > \rho(u)$ and $u \in C^{v}(v_1)$.
\end{lemma}

\begin{proof}
    The fact that the ordered $\Theta$-routing step implies the two conditions follows by definition. Hence, we focus on the other proof direction. Suppose $v \in N(u)$ satisfies the two conditions. Then, since $\rho(v) > \rho(u)$, the edge $e = \{u,v\} \in E$ must have been created upon the insertion of $v$ into $G$ (when we inserted $u$ into $G$, $v$ did not exist yet). This implies that, when $v$ was inserted into $G$, $u$ was the closest vertex in $C^{v}(v_1)$. It follows that there does not exist a vertex $u' \ne u$ such that $u' \in \Delta{vu} \subset C^{v}(v_1)$ and $\rho(u') < \rho(v)$, because the existence of such a vertex $u'$ contradicts the fact that $u$ was the closest vertex in $C^{v}(v_1)$ when $v$ was inserted into $G$. Hence, an ordered $\Theta$-routing step at $v$ towards $v_1$ leads to $u$.
\end{proof}

Note that checking the \emph{implication} of the two conditions $\rho(v) > \rho(u)$ and $u \in C^{v}(v_1)$, i.e., that an ordered $\Theta$-routing step at $v$ towards $v_1$ leads to $u$, would require $2$-local information -- given a current vertex $u$, accessing a neighbor $v \in N(u)$ is $1$-local, and accessing a neighbor $v' \in N(v)$, in order to check whether an ordered $\Theta$-routing step at $v$ towards $v_1$ leads to $v'$ such that $v' = u$, is $2$-local. So, instead, $\mathcal{A}^{up}$ checks the two conditions, which is $1$-local -- it just compares the order of $u$, $v$, and $t$, and checks the geometric statement $u \in C^{v}(v_1)$, both of which are $1$-local.

Next, we look at the relation between $\mathcal{A}^{up}$ on $G$ and a depth-first search on $T$.

\begin{lemma} \label{lmm:DFS-on-T}
    Algorithm $\mathcal{A}^{up}(v_1, t)$ on $G$ uses an edge $\{u,v\}$ if and only if a depth-first search from $v_1$ on $T$ uses $\{u,v\}$. 
\end{lemma}

\begin{proof}
    When $\mathcal{A}^{up}$ takes an edge $\{u,v\}$, it is either an exploration or a backtracking step. Since each backtracking step takes an edge of $T$ by definition, \cref{lmm:two-conditions} concerning exploration steps suffices to conclude that $\mathcal{A}^{up}$ only takes edges of $T$, as the algorithm explicitly checks the two conditions stated in the lemma to determine whether an edge should be taken as an exploration step.

    When $\mathcal{A}^{up}$ explores a neighbor $v$ of $u$, by~\cref{lmm:ordered-theta-routing-towards-v1} we implicitly ``know'' the current exploration path from $v_1$ to $v$, as it can be traversed in reverse from $v$ to $v_1$ using ordered $\Theta$-routing. 

    If $\mathcal{A}^{up}$ backtracks from a vertex $v$ to $u$, then $v$ is stored in $u_p$ and thus the counter-clockwise sweeping of half-line $l$ at $u$ resumes to identify the next unexplored ordered $\Theta$-routing edge pointing to $u$ if it exists. By~\cref{lmm:two-conditions} and the counter-clockwise sweeping of $l$, each ordered $\Theta$-routing edge pointing to $u$, which is an edge in $T$, is explored at most once. Consequently, $\mathcal{A}^{up}$ explores any neighbor $v$ of $u$ at most once and backtracks from $v$ to $u$ at most once.
\end{proof}

Since $T$ is a spanning tree of $G$ by~\cref{lmm:spanning-tree} and algorithm $\mathcal{A}^{up}(v_1, t)$ performs a depth-first search on $T$ by~\cref{lmm:DFS-on-T}, the algorithm routes from $v_1$ to $t$ in $O(n)$ hops.

\begin{corollary} \label{cor:reach-t-in-linear-hops}
    $\mathcal{A}^{up}(v_1, t)$ reaches $t$ in $O(n)$ hops.
\end{corollary}

Next, we show that $\mathcal{A}^{up}(v_1, t)$ performs a linear number of operations to determine all neighbors to be explored.

\begin{lemma} \label{lmm:linear-number-of-candidate-verifications}
    Given any current vertex $u$ visited by $\mathcal{A}^{up}(v_1, t)$, let a \emph{candidate verification} of $v \in N(u)$ be the operations required to check two conditions $\rho(v) > \rho(u)$ and $u \in C^{v}(v_1)$. Then, $\mathcal{A}^{up}(v_1, t)$ performs at most $O(kn)$ candidate verifications before reaching $t$, where $n = |V|$ and $k$ is the number of cones of $G = (V,E)$.
\end{lemma}

\begin{proof}
    We know that algorithm $\mathcal{A}^{up}(v_1, t)$ explores each vertex in $V$ at most once by~\cref{lmm:DFS-on-T}. Since there is only a single sweep of $l$ to check the neighbors of the current vertex, each edge in $E$ contributes to the number of candidate verifications at most twice, once at each endpoint. So, the number of candidate verifications is upper bounded by $2|E| < 2 kn$.
\end{proof}

We observe that, to reach $t$, we actually only need to consider vertices whose order is smaller than $\rho(t)$ during the exploration due to the order-monotonicity of the ordered $\Theta$-routing path from $t$ to $v_1$, which informs a simple modification to algorithm $\mathcal{A}^{up}$: let it check $\rho(v) \leq \rho(t)$ in addition to $\rho(v) > \rho(u)$ and $u \in C^{v}(v_1)$. This makes the algorithm follow the same edges as a depth-first search on a subtree $T' = T[V']$ of $T$ induced by $V' \subseteq V$ such that $V'$ contains all vertices with order at most $\rho(t)$. However, we observe that the asymptotic complexity of $\mathcal{A}^{up}$ remains the same after this modification, so we keep the original $\mathcal{A}^{up}$ for simplicity.

We also observe that the counter-clockwise order of exploration is arbitrary. We can order them in a different way as long as the order of exploring neighbors is well-defined and we can determine locally where to continue our exploration after backtracking, without storing anything about the previously explored neighbors. For example, we can explore neighbors $v$ of $u$ in increasing or decreasing order of $\rho(v)$ or in increasing or decreasing order of $|uv|$. We implicitly assume that the exploration ordering of the neighbors is consistent with the order in which the neighbors are stored during the construction of the graph. This ensures that, after a backtracking step, we can efficiently obtain the next neighbor to explore. 

\subsubsection{Routing from $s$ to $t$}

As $\mathcal{A}^{down}$ routes from $s$ to $v_1$ and $\mathcal{A}^{up}$ routes from $v_1$ to $t$, we obtain our final algorithm that routes from $s$ to $t$.

\begin{algorithm}[$\mathcal{A}(s, t)$]
    Given the source vertex $s \in V$ and the destination $t \in V$, the algorithm routes from $s$ to $t$ by performing the following two steps:
    \begin{enumerate}
        \item Execute $\mathcal{A}^{down}(s)$ to route from $s$ to $v_1$.
        \item Execute $\mathcal{A}^{up}(v_1, t)$ to route from $v_1$ to $t$.
    \end{enumerate}
\end{algorithm}

Consequently, we obtain the following theorem.

\thmAlgorithmA*

\begin{proof}
    By~\cref{lmm:routing-to-the-first-inserted-vertex}, the local routing algorithm $\mathcal{A}^{down}(s)$ initiated at $s$ takes $O(n)$ hops and $O(kn)$ time in the worst-case to reach $v_1$. By~\cref{cor:reach-t-in-linear-hops}~and~\cref{lmm:linear-number-of-candidate-verifications}, the local routing algorithm $\mathcal{A}^{up}(v_1, t)$ requires at most $O(n)$ hops and $O(kn)$ candidate verifications (and thus $O(kn)$ time overall) to route from $v_1$ to $t$. Therefore, $\mathcal{A}(s, t)$ takes $O(n)$ hops and $O(kn)$ time to route from $s$ to $t$.

    In addition, $\mathcal{A}(s, t)$ is a deterministic $O(1)$-memory $1$-local routing algorithm, because $\mathcal{A}^{down}$ is a deterministic memoryless $1$-local routing algorithm by~\cref{lmm:routing-to-the-first-inserted-vertex}, $\mathcal{A}^{up}$ is a deterministic $O(1)$-memory $1$-local routing algorithm by~\cref{lmm:deterministic-constant-memory-1-local}, and we need only a single additional bit to store in which of the two phases the algorithm is.
\end{proof}

\section{Conclusion}

In this paper, we addressed the following problem: \emph{Does there exist a local routing algorithm that works on the ordered $\Theta$-graph?} We first proved there does not exist a deterministic memoryless $h$-local routing algorithm on the ordered $\Theta_{k}$-graph for any integers $h \geq 1$ and $k \geq 2$. Then, we presented a deterministic $O(1)$-memory local routing algorithm that is guaranteed to reach the destination in $O(n)$ hops and $O(kn)$ time on the ordered $\Theta_{k}$-graph for any $k \geq 2$. To the best of our knowledge, the algorithm we provided is the first local routing strategy that is successful on the ordered $\Theta_{k}$-graph.

Although our local routing algorithm $\mathcal{A}(s,t)$ can route between any source and destination, we remark that the total routing path length can be arbitrarily high, which means our local routing algorithm is not $c$-competitive.

Therefore, one natural open question is whether there exists a $c$-competitive local routing algorithm using a small amount of memory on the ordered $\Theta$-graph. Things that make finding such an algorithm difficult include the fact that ordered $\Theta$-graphs are generally not plane, the proofs used to bound the spanning ratio by Bose et al.~\cite{Bose-etal:2004:OrderedThetaGraphs} and Bose et al.~\cite{Bose-etal:2016:ThePriceOfOrder} are highly non-local, and the commonly used properties (such as empty canonical triangles) do not hold for these graphs.

\bibliography{paper.bib}
\end{document}